\documentclass[preprint,amsmath]{revtex4}
\input {epsf.sty}
\usepackage{epsfig}
\usepackage{amsmath,amsthm,amssymb}

\begin{document}
\title{Nano-Torsional Resonator Torque Magnetometry}
\author{J.P. Davis$^1$, D. Vick$^2$, D.C. Fortin$^1$, J.A.J. Burgess$^{1,2}$, W.K. Hiebert$^2$ and M.R. Freeman$^{1,2}$}
\affiliation{$^{1}$ Department of Physics, University of Alberta, Edmonton, Alberta, Canada T6G 2G7}
\affiliation{$^{2}$ National Institute for Nanotechnology, Edmonton, Alberta, Canada T6G 2M9}

\date{Version \today}

\begin{abstract}  Magnetic torque is used to actuate nano-torsional resonators, which are fabricated by focused-ion-beam milling of permalloy coated silicon nitride membranes.  Optical interferometry is used to measure the mechanical response of two torsion modes at resonance, which is proportional to the magnetization vector of the nanomagnetic volume.  By varying the bias magnetic field, the magnetic behavior can be measured with excellent sensitivity ($\approx 10^8 \mu_B$) for single magnetic elements.
\end{abstract}

\maketitle

Nanomechanical resonators have proven their capabilities for both mass and force sensing \cite{Rug04,Li07,Nai09}.  In terms of mechanical sensing of torque, or moment of inertia, there has been a progression from sensitive macroscale torsion balances \cite{Bar12} to microscale resonators \cite{Cha05}, but nanoscale torsional resonators \cite{Cle98, Moh02} are necessary to push the limits of sensitivity for mechanical torque measurements \cite{Zol08}.  Here we report the use of magnetic torque to drive torsion modes of permalloy coated nanoresonators and show that the interferometric response can be used to measure their magnetic behavior.  This is a sensitive technique for studying the magnetism of single nanoscale magnetic objects without the need for significant averaging \cite{All03, Zhi08}.  The simplicity and sensitivity of this technique will soon make nanomechanical torque magnetometry a common implement in the toolbox of the nanomagnetism researcher.  

Torque magnetometry has been used for many years \cite{Wil37} for a variety of magnetic measurements, from the de Hass-van Alphen effect \cite{Con63}, to vortices in superconductors \cite {Li07b}, and can be performed in fields that are too high for many other sensitive magnetic techniques \cite{Col09, Ble09}.  The concept behind mechanical torque magnetometry is straightforward, with a number of variations.  For example, one can drive a torsional resonator at its resonance frequency using non-magnetic actuation.  In this scenario, magnetic material on the resonator experiences a torque in the presence of an external magnetic field.  This in turn either shifts the resonance frequency \cite{Har99}, or produces dissipation altering the mechanical $Q$ \cite{Sti01}.  Magnetic resonance force microscopy is another impressive form of torque magnetometry \cite{Obu08}.  In the variation used in this letter, the magnetic torque is used to drive the mechanical modes at resonance.  In this way, the amplitude of the mechanical displacement originates solely from the magnetization.

\begin{figure}[b]
%%%%%%%%%%%%%%%%%   F I G U R E  1   %%%%%%%%%%%%%%%%%%
\centerline{\includegraphics[width=2.8in]{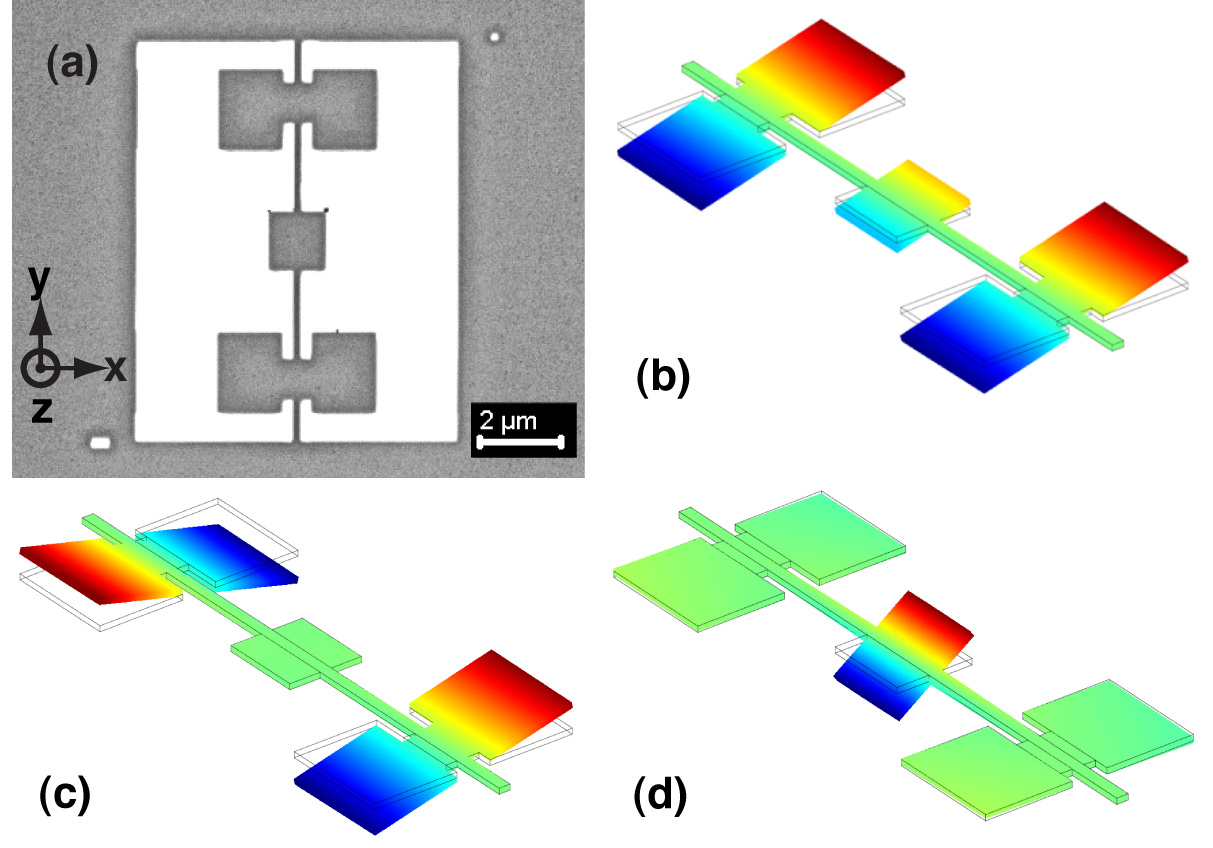}}
%%%%%%%%%%%%%%%%%%%%%%%%%%%%%%%%%%%%%%%%%%%%%
\caption{{\footnotesize\label{fig1} Color online.  Nano-torsional resonators: (a) SEM micrograph, the permalloy is on the opposite side.  The Cartesian coordinate system is indicated.  (b) Finite element model of first torsional mode.  (c) The second torsional mode.  (d) The third torsional mode.  In (b)-(d) the color scale indicates the relative amplitude of z-displacement.}}
\end{figure}
	
\begin{figure}[t]
%%%%%%%%%%%%%%%%%   F I G U R E  2   %%%%%%%%%%%%%%%%%%
\centerline{\includegraphics[width=2.8in]{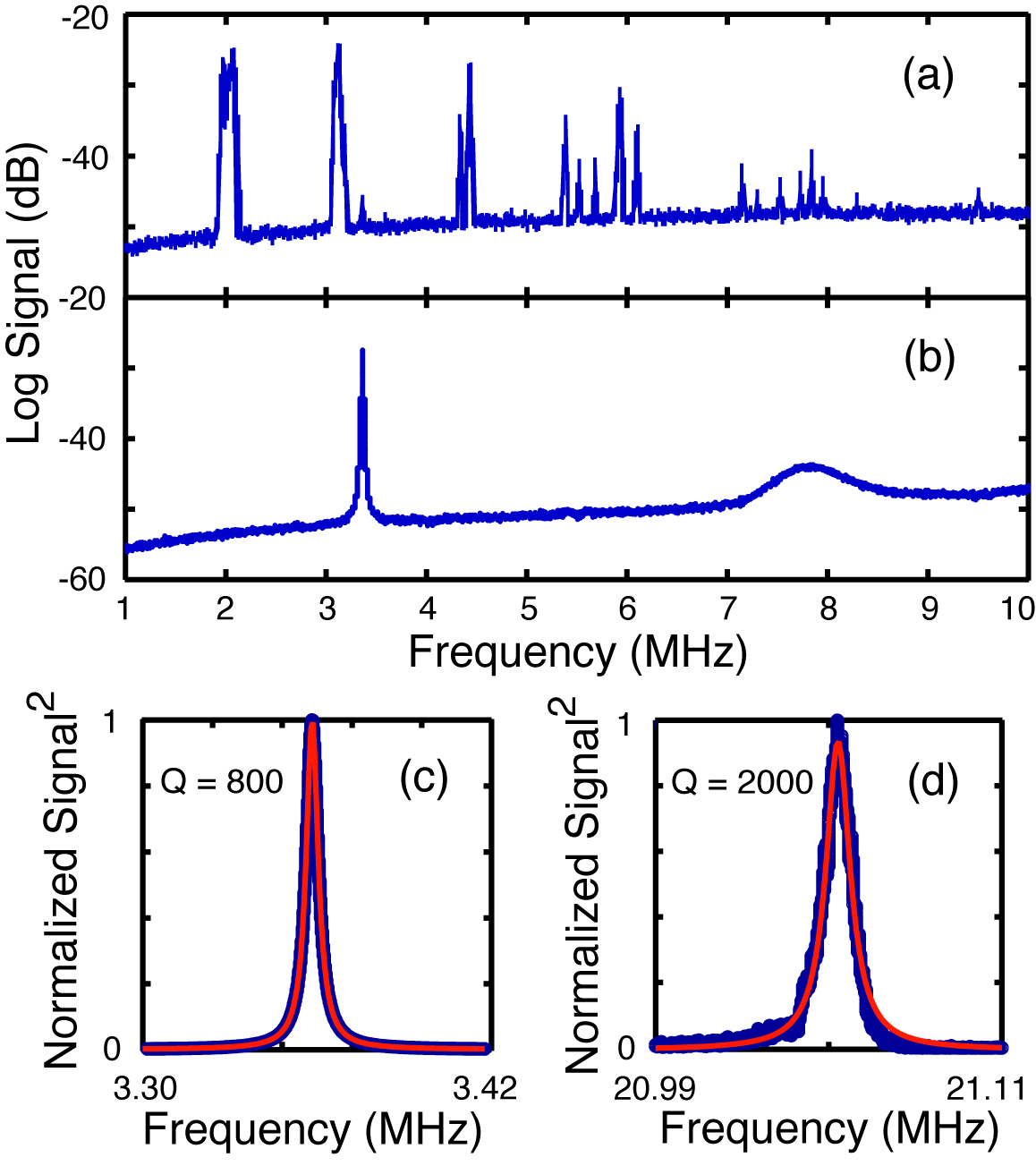}}
%%%%%%%%%%%%%%%%%%%%%%%%%%%%%%%%%%%%%%%%%%%%%
\caption{{\footnotesize\label{fig2} Color online.  Optical interference spectrum of a nano-torsional resonator: (a) Using broadband piezo actuation, showing numerous mechanical modes.  (b) Using magnetic actuation: the magnetic torque selectively drives the torsion mode.  The broad peak below 8 MHz is electrical crosstalk.  (c) The square of the linear spectrum for the first torsional mode, with a Lorenztian fit, which gives a $Q$ of 800, and (d) the same for the third torsional mode which has a $Q$ of 2000.  The SNR for this mode is smaller than for the first torsional mode.}}
\end{figure}
	
The sensitivity of a torsional resonator is inversely proportional to its torsion constant, $\kappa$, which decreases with its cross-section.  Specifically, the static displacement of a torsion paddle is $z = (w/2)\sin \theta$, where $w$ is the full width of the paddle, $z$ is out of the plane of the resonator, and the angle of deflection of the torsion rod from an external torque, $\tau$, is $\theta = \tau / \kappa$.  The torsion constant for a bar of rectangular cross section is $\kappa = \beta d t^3E/2l(1+\nu)$, where $l$ is the length of the torsion rod, $d$ is its width, $t$ is its thickness, $\beta$ is a geometrical factor approximately $0.2$, $E$ is the Young's modulus and $\nu$ is the Poisson's ratio \cite{Cle04}.  Reducing the dimensions of the torsion rod to the nanoscale ($d=200$~nm and $t=100$~nm) decreases the torsion constant and increases the sensitivity of the torsional resonator.  The torsional spring constant of our resonator is $\approx 5\times10^{-13}$~N$\cdot$m, which is roughly three orders of magnitude smaller than previous micromechanical torque magnetometers \cite{Cha05}.  The torque on the resonator from the external drive field, $H_z$, is $\boldsymbol{\tau} = \boldsymbol{\mu} \times \boldsymbol{H_z} = \boldsymbol{m} V \times \boldsymbol{H_z}$, where $\boldsymbol m$ is the magnetic moment per unit volume, $V$, in the $x$-direction.  The detection sensitivity can be expressed as $m_s V$, the saturation magnetic moment, divided by the signal to noise ratio (SNR).  This gives $\approx 10^8 \mu_B$ for a single magnetic field sweep for our resonators.  This is better than a single field sweep for traditional magnetic detection techniques, for example an order of magnitude better than magneto-optical Kerr microscopy \cite{All03}.  

Resonators are fabricated using focused-ion-beam milling of commercial low-stress silicon nitride membranes \cite{Nor}, Fig.~1a.  Silicon nitride has been shown to posses excellent mechanical properties \cite{Zwi08,Ver08}.  The thin silicon nitride results in a thin torsion rod, important for keeping a low torsion constant and a sensitive resonator, as discussed above.  Here we use 100 nm thick membranes, coated on one side with 10 nm of permalloy and on the other with a thin layer of gold ($\approx 2$~nm) to prevent charging during milling.  The permalloy is deposited in UHV ($\approx 1\times10^{-9}$ torr during deposition) using a collimated source.  The gold side faces the ion-beam during milling, so that the nitride thickness prevents ion damage from affecting the permalloy.

The triple paddle design that we use here is better suited to the thin silicon nitride membranes than the traditional double paddle resonator design \cite{Kle85}.  By having two clamped ends we do not sacrifice the structural integrity of the resonator, while maintaining most of the mechanical advantages of the double paddle resonator.  The triple paddle design results in a simple set of torsional modes, of which we take advantage.  Finite element models of these modes are shown in Figs.~1b-d.  In the first torsional mode, Fig.~1b, all three paddles oscillate in phase, with a resonant frequency of 3.36 MHz.  The second torsional mode, Fig.~1c, is the antisymmetric mode, where the two large paddles oscillate out of phase with one another and the center paddle is essentially stationary.  This mode is not magnetically actuated in the current configuration.  The third torsional mode, Fig.~1d, is an oscillation primarily of the center paddle alone, and has a resonant frequency of 21 MHz.  This mode is the most isolated from the surrounding membrane and therefore can exhibit a higher mechanical $Q$.  A feature of the triple paddle design is that the resonance frequencies of the various modes can be semi-independently tuned by changing the geometry of the relevant portion of the resonator.

Detection is performed using optical interference.  The silicon nitride membrane is mounted $\approx 30~\mu$m from a smooth silicon wafer.  This serves as one mirror of a crude Fabry-Perot cavity, with the resonator itself as the other mirror.  This assembly is mounted on a broadband piezoelectric transducer for mechanical actuation, and placed in front of a two-axis Hall probe inside of an optical access UHV chamber.  The measurements reported here were taken at $\approx 2\times10^{-9}$ torr.  An excitation coil is mounted near the sample for the torquing magnetic field, $H_z$, and the applied bias field, $H_0$, is generated by a permanent magnet that can be rotated or retracted.  The maximum torque is applied when a strong $H_0$ is applied perpendicular to the torsion rod, along the $x$-axis in Fig.~1a.  The broadband piezoelectric, or the $H_z$ coil, is driven using an amplified tracking signal from a spectrum analyzer (HP 8594E with option 010) and the optical signal is detected using a photomultiplier tube, amplified, and sent to the spectrum analyzer.  For magnetic hysteresis measurements the signal is detected at a fixed frequency using an RF lock-in amplifier (EG\&G 5202).
	
\begin{figure}[b]
%%%%%%%%%%%%%%%%%   F I G U R E  3   %%%%%%%%%%%%%%%%%%
\centerline{\includegraphics[width=3.3in]{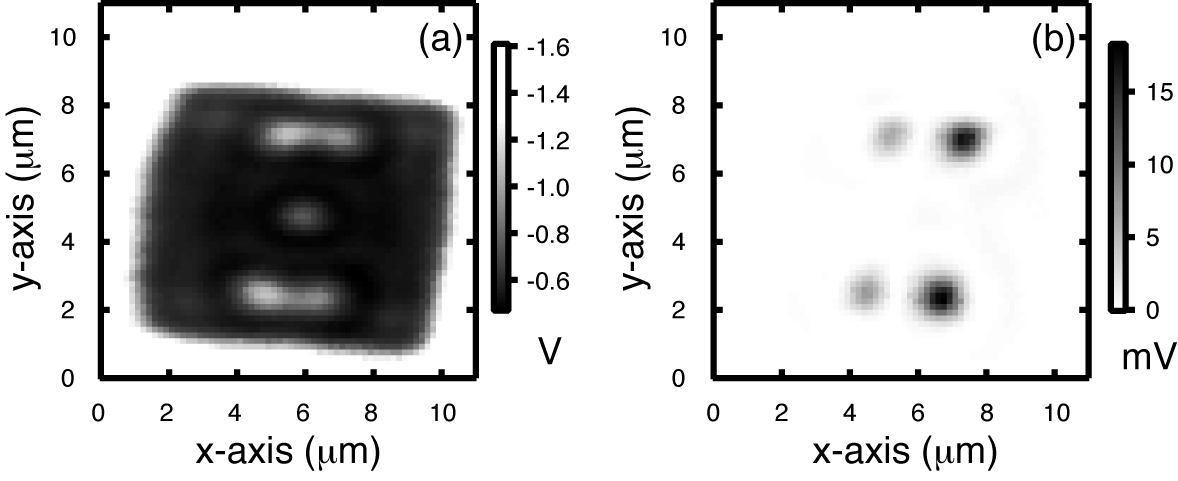}}
%%%%%%%%%%%%%%%%%%%%%%%%%%%%%%%%%%%%%%%%%%%%%
\caption{{\footnotesize\label{fig3} Raster scanned optical image.  (a) Reflected intensity.  (b) Simultaneous measurement of the interferometric signal at the frequency of the first torsional mode, showing that it is localized to the resonator.  Asymmetry is a result of non-parallelism of the Fabry-Perot cavity.}}
\end{figure}
	
The detection laser (632 nm) is focused onto the resonator using a long working length objective through a window in the UHV chamber and 300 nW of optical power strikes the sample.  At this optical power there are no heating effects on the resonator, the photomultiplier tube is not saturated and the measurement is shot-noise limited.  The optics are in a cage system affixed to a scan stage with piezo control.  This enables spatial scanning of the focused spot, allowing detection of optical interferometric signal at precise locations on the resonator and also to ensure that the resonant signal is localized to the nano-torsional resonator.
				
\begin{figure}[t]
%%%%%%%%%%%%%%%%%   F I G U R E  4   %%%%%%%%%%%%%%%%%%
\centerline{\includegraphics[width=3.0in]{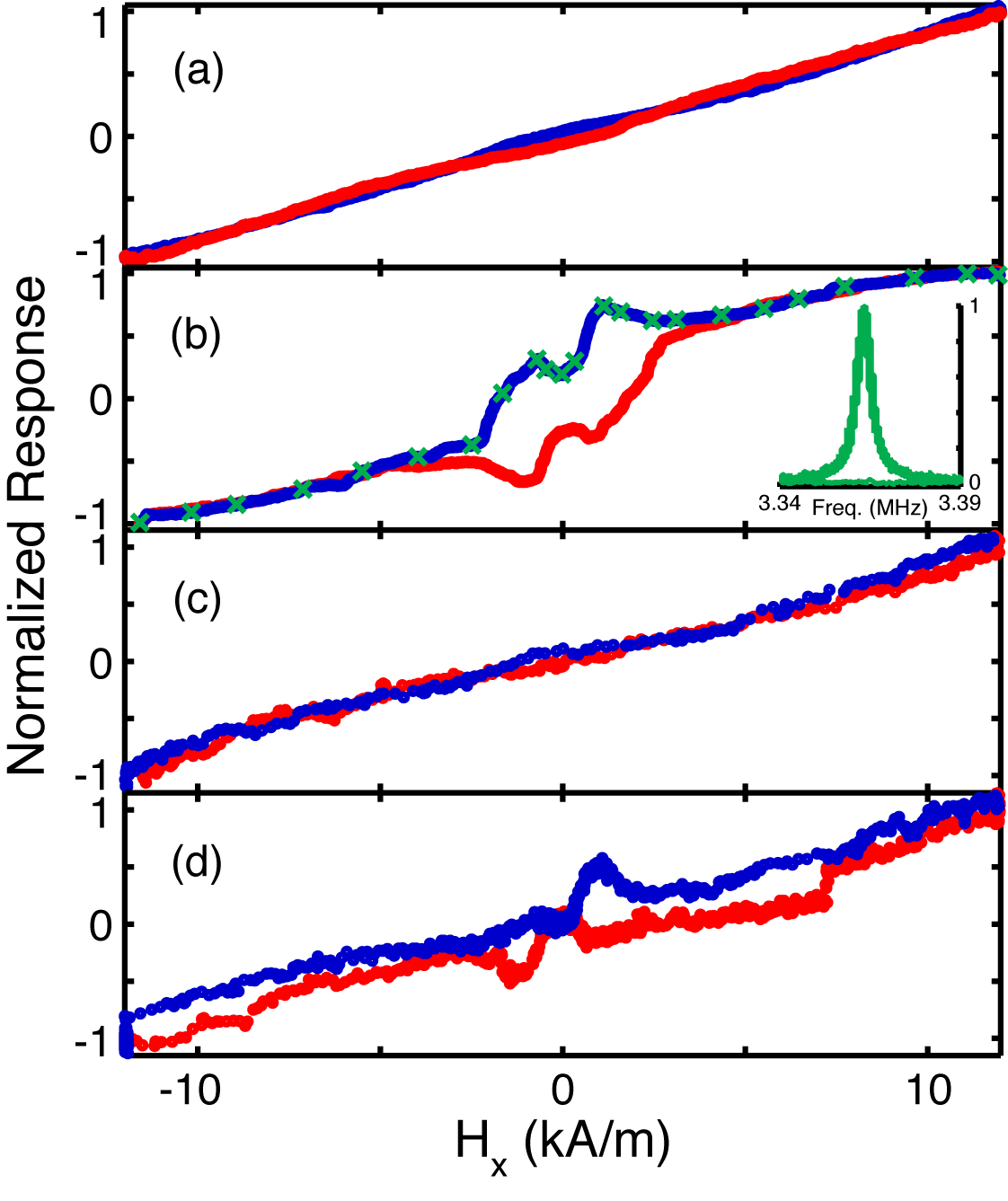}}
%%%%%%%%%%%%%%%%%%%%%%%%%%%%%%%%%%%%%%%%%%%%%
\caption{{\footnotesize\label{fig4}
Color online.  Interferometric signals, proportional to $M_x$, from the first and third torsion modes as a function of the applied magnetic field, $H_x$.  Red curves are increasing fields and blue are decreasing.  (a-b) The first torsion mode.  (a) The applied field ($H_0 = 12$ kA/m) was rotated in the plane of the resonator and averaged over 10 runs.  (b) $H_x$ was swept, while $H_y =0$, and averaged over 3 runs.  Inset to (b) shows twenty three frequency sweeps at the magnetic fields marked with green X's in (b), all normalized.  (c-d) Single field sweeps of the third torsion mode at the center paddle.  In (c) the applied field was rotated ($H_0 = 12$ kA/m) in the plane of the resonator and in (d) $H_x$ was swept, while $H_y =0$.}}
\end{figure}
	
In Fig.~2, we show resonance spectra for the nano-torsional resonator.  The spectrum for the piezo actuation, Fig.~2a, is complex and many of the peaks are coupled modes between the resonator and the surrounding membrane.  The first torsional resonance  (at 3.36 MHz) is barely visible, as the piezo preferentially drives flexural modes.  The corresponding magnetic actuation spectrum is considerably simpler, Fig.~2b, since the torsional modes are selectively driven.  Figure 3 shows a spatial scan of the reflected intensity and a simultaneous measurement of the amplitude of the resonance peak of the first torsional mode, verifying that the magnetically actuated mode is localized to the paddles.  The mechanical $Q$'s are 800 and 2000 for the first and third torsion modes respectively, Fig.~2c and 2d.  The detection sensitivity of the third torsional mode is decreased compared to the first for two reasons: the torque on the resonator is significantly smaller because of the smaller volume of magnetic material coupled to this mode, and the displacement amplitude is smaller for the center paddle.

The nano-torsional resonators can be used to study the magnetic behavior of nanoscale magnetic elements.  In Figs.~4a-b we show the peak amplitude of interferometric response for the first torsional mode, which is sensitive to all of the magnetic material on the resonator, as a function of applied magnetic field, $H_x$.  In Fig.~4a, a fixed-strength field is rotated in the plane of the resonator.  In this way, domain switching is avoided and instead the magnetization follows the bias field.  The interferometric response is expected to be nearly linear in $H_x$, with small oscillations in $M_x$ that are out of phase between the increasing and decreasing field sweeps as result of the demagnetization field shape anisotropy \cite{Xue06}.  This is what is observed.  

In Fig.~4b we show the interferometric response for field sweeps in which the bias field $H_x$ is varied and $H_y = 0$.  There are two sections of $M_x$, in each sweep direction, in which the magnetic response increases, likely due to switching involving the film along the torsion rod.  We have tested to ensure that these are not a result of frequency shifts or dissipation changes.  In the inset of Fig.~4b we show twenty three frequency sweeps at various points in Fig.~4b, all normalized.  The fact that they collapse to a universal curve when normalized verifies that the response to the magnetic field is a result of the varying $x$-component of the magnetization of the sample.  The domain structure for the permalloy film on the triple paddle resonator is complex and difficult to interpret without thorough micromagnetic simulations.  

Figs.~4c-d show field sweeps for the third torsional mode, which is sensitive to the magnetic material on the center paddle.  These are \textit{single} sweeps, with no smoothing or averaging, demonstrating the sensitivity of this technique.  In Fig.~4c, similar oscillations in the magnetization can be seen as in Fig. 4a, arising from demagnetization field shape anisotropy.  Likewise, the  magnetization behavior in Fig.~4d is similar to that of Fig.~4b.  The magnetization in Fig.~4b is closer to saturation at 12 kA/m than in Fig.~4d, as expected for a larger magnetic area.  We note that there are a few steps in Fig.~4d, most obviously at 7.2 kA/m on the increasing field sweep.  This is a Barkhausen step, which originates from the discreet nature of magnetic domain switching \cite{Bar19}, indicating that the nano-torsional technique is sensitive enough to resolve single domain events in nanoscale magnetic elements.

In conclusion, we have shown that by fabricating torsional resonators with nanoscale torsion rods one can achieve extremely low torsion constants, which translates into sensitivity for torque magnetometry.  Our nano-torsional resonators are straightforward to fabricate, using commercial silicon nitride membranes that have been coated with magnetic material followed by focused-ion-beam milling.  Magnetic actuation selectively drives the torsional modes and is directly proportional to the magnetization vector of the magnetic elements, enabling sensitive studies of nanomagnetic elements.  One figure of merit for mechanical torque magnetometry is the minimum detectable spin signal for a thermally limited measurement and a measurement bandwidth of 1 Hz: $\mu_{min} = \sqrt{4k_BT \kappa/Q\omega_0}/H_z$.  With our maximum drive field, $H_z = 800$ A/m, that that leads to a minimum detectable signal of $2\times10^5\mu_B$ for the first torsional mode and $6\times10^4\mu_B$ for the third, which corresponds to a cube of permalloy 9 nm on a side.  Our detection sensitivity is not thermally limited because of the small scale of thermomechanical displacement.  Recently, there have been dramatic advances in displacement detection by coupling high finesse optical cavities to nanomechanical resonators allowing thermomechanical displacement of nanoscale resonators to be measured \cite{Ane09}.  
Improving our detection sensitivity to be thermally limited will allow nano-torsional resonator torque magnetometry to rival any proposal for magnetic detection \cite{Rod09} and open up unknown regimes of nanoscale magnetic measurements.

We acknowledge support from NSERC, iCORE, CIFAR, NINT, CRC, AIF, and the Integrated Nanosystems Research Facility supported through CFI and nanoAlberta.  We would like to thank Don Mullin, Greg Popowich and Tony Walford for their contributions.


\begin{thebibliography}{xxx}

{\footnotesize

\bibitem{Rug04}
D. Rugar, R. Budakian, H.J. Mamin and B.W. Chul, Nature \textbf{430}, 329 (2004).

\bibitem{Li07}
M. Li, H.X. Tang and M.L. Roukes, Nature Nanotech. \textbf{2}, 114 (2007).

\bibitem{Nai09}
A.K. Naik, M.S. Hanay, W.K. Hiebert, X.L. Feng and M.L. Roukes, Nature Nanotech. 
\textbf{4}, 445 (2009).

\bibitem{Bar12}
G. Barlow, Proc. of the Royal Society of London, \textbf{87}, 1 (1912).

\bibitem{Cha05}
M.D. Chabot, J.M. Moreland, L. Gao, S.H. Liou and C.W. Miller, Journal of MEMS \textbf{14}, 1118, (2005).

\bibitem{Cle98}
A.N. Cleland and M.L. Roukes, Nature \textbf{392}, 160 (1998).

\bibitem{Moh02}
P. Mohanty, D.A. Harrington, K.L. Ekinci, Y.T. Yang, M.J. Murphy and M.L. Roukes, Phys. Rev. B \textbf{66}, 085416 (2002).

\bibitem{Zol08}
G. Zolfagharkani, A. Gaidarzhy, P. Degiovanni, S. Kettemann, P. Fulde and P. Mohanty, Nature Nanotech. \textbf{3}, 720 (2008).

\bibitem{All03}
D.A. Allwood, G. Xiong, M.D. Cooke and R.P. Cowburn, J. Phys. D: Appl. Phys. \textbf{36}, 2175 (2003).

\bibitem{Zhi08}
Z. Liu, R.D. Sydora and M.R. Freeman, Phys. Rev B \textbf{77}, 174410 (2008).

\bibitem{Wil37}
H.J. Williams, Rev. Sci. Inst. \textbf{8}, 56 (1937).

\bibitem{Con63}
J.H. Condon and J.A. Marcus, Phys. Rev. \textbf{134}, A446 (1963).

\bibitem{Li07b}
L. Li, J.G. Checkelsky, S. Komiya, Y. Ando and N.P. Ong, Nature Phys. \textbf{3}, 311 (2007).

\bibitem{Col09}
A.I. Coldea, C.M.J. Andrew, J.G. Analytis, R.D. McDonald, A.F. Bangura, J.-H. Chu, I.R. Fisher and A. Carrington, Phys. Rev. Lett. \textbf{103}, 026404 (2009).

\bibitem{Ble09}
A.C. Bleszynski-Jayich, W.E. Shanks, B. Peaudecerf, E. Ginossar, F. von Oppen, L. Glazman and J.G.E. Harris, Science \textbf{326}, 272 (2009).

\bibitem{Har99}
J.G.E. Harris, D.D. Awschalom, F. Matsukura, H. Ohno, K.D. Maranowski and A.C. Gossard, App. Phys. Lett. \textbf{75}, 1140 (1999).

\bibitem{Sti01}
B.C. Stipe, H.J. Mamin, T.D. Stowe, T.W. Kenny and D. Rugar, Phys. Rev. Lett. \textbf{86}, 2874 (2001).

\bibitem{Obu08}
Yu. Obukhov, D.V. Pelekhov, J. Kim, P. Banerjee, I. Martin, E. Nazaretski, R. Movshovich, S. An, T.J. Gramila, S. Batra and P.C. Hammel, Phys. Rev. Lett. \textbf{100}, 197601 (2008).

\bibitem{Cle04}
A.N. Cleland, ``Foundations of Nanomechanics'' (Springer, Berlin 2004).

\bibitem{Nor}
Norcada, 4465 - 99 St. Edmonton, Alberta, Canada T6E 5B6.	

\bibitem{Zwi08}
B.M. Zwickl, W.E. Shanks, A.M. Jayich, C. Yang, A.C. Bleszynski Jayich, J.D. Thompson and J.G.E. Harris, Appl. Phys. Lett. \textbf{92}, 103125 (2008). 

\bibitem{Ver08}
S.S. Verbridge, H.G. Craighead and J.M. Parpia, Appl. Phys. Lett. \textbf{92}, 013112 (2008).

\bibitem{Kle85}
R.N. Kleiman, G.K. Kaminsky, J.D. Reppy, R. Pindak and D.J. Bishop, Rev. Sci. Instrum. \textbf{56}, 2088 (1985).

\bibitem{Xue06}
D. Xue, X. Fan and C. Jiang, Appl. Phys. Lett. \textbf{89}, 011910 (2006).

\bibitem{Bar19}
H. Barkhausen, Physikalische Zeitschrift, \textbf{20}, 401 (1919).

\bibitem{Ane09}
G. Anetsberger, O. Arcizet, Q.P. Unterreithmeier, R. Rivi\`ere, A. Schliesser, E.M. Weig, J.P. Kotthaus and T.J. Kippenberg, Nature Phys. \textbf{5}, 909 (2009).

\bibitem{Rod09}
I. Rod, O. Kazakova, D.C Cox, M. Spasova and M. Farle, Nanotechnology \textbf{20}, 335301 (2009).
}

\end{thebibliography}
\end{document}